\documentclass{article}
\begin{document}
\date{}
\title{Cosmic Intelligence and Black Holes}
\author{Vladimir A. Lefebvre\\School of Social Sciences\\University of California,
Irvine\\and\\Yuri N. Efremov\\Sternberg Astronomical Institute\\Moscow State University,
Moscow}
\maketitle

\begin{abstract}

The paper is devoted to a new direction in SETI. After a general discussion of the field, the
authors put forth the hypothesis that the black holes may serve as a physical substratum for
intelligent beings. This hypothesis is based on four parallels between the brain-psyche system, on
the one hand, and black holes, on the other. (1) The descriptions of brain and psyche, in the
system brain-psyche, are complementary to each other, as descriptions by internal and external
observers of a black hole in Susskind-t'Hooft's schema. (2) There is an aspect of the inner
structure of a black hole in Kerr's model of the rotating black hole that is isomorphic to the
structure of the human subjective domain in the psychological model of reflexion. (3) Both black
holes and the brain-psyche system have a facet which can be represented using thermodynamic
concepts. (4) The brain lends itself to a holographic description; as has been recently
demonstrated by Susskind, black holes can also be described holographically. The authors
speculate that the intelligent black holes can generate other black holes by triggering creation of
massive stars from which new black holes arise after the stars' collapse. In addition, the authors
analyze certain strange phenomena related to the birth of young massive stars which may or may
not be connected with such a triggering.

\end{abstract}

\section{Introduction. The Search for Extraterrestrial Civilizations}

This article was written by two men no longer young. They first met a half-century ago in the
school Astronomy Club of the Moscow Planetarium. Subsequently one of them became a
professional astronomer, the other a psychologist and anthropologist. Like the many other young
astronomers gathering every week in the Planetarium's Star Room, they were fascinated, among
the other issues of this science, by the question of whether or not we are alone in the Universe.
This question remains with us to this day, along with the memory of the stars' twinkling in the
planetarium dome and the whole festive atmosphere of our first contact with knowledge. It was
seemingly no accident that our fellow member N.S. Kardashev became one of the pioneers in the
search for extraterrestrial civilizations.

Unfortunately, this problem has yet to acquire serious scientific status. One of the reasons for this
is that the devotees of this field formulate their goals guided only by intuition and common sense.
The proposal of hypotheses regarding the types of activity of extraterrestrial intelligence, in
essence, is little different from the free inventions of science fiction writers. The search for
cosmic civilizations will acquire the status of a strictly scientific undertaking only if it becomes
possible to construct a theoretical model of the world, one of whose natural components would
be the intelligent subject. Such a model should connect mind as a phenomenon with our physical
understanding of the Universe, to shed light on the nature of the physical objects we seek to find. 

Such an undertaking is fully suited to the spirit of the natural sciences at the turn of the new
century. We are the witnesses of a new scientific revolution in physics comparable with the one
whose results let to the appearance of quantum mechanics. The unification of quantum
mechanics with the theory of relativity is taking place before our eyes. A storm of research is
going on in two fields, whose code names are "quantum gravitation" and "string theory." In this
article we attempt to conduct a series of theoretical and empirical arguments that this scientific
revolution may culminate in something unexpected -- the inclusion of the intelligent subject in a
unified purview. And it is precisely such a unified model that may indicate to us the objective
traits of intelligent subjects.

\section{The Silence of the Universe}

Today we possess a detailed view of the Universe, and nowhere do we find barriers to the spatial
propagation of intelligent life, once it has appeared. Nonetheless, all attempts to identify
extraterrestrial intelligence have so far been futile. Enrico Fermi's well-known question "Where
are THEY?" turned the absence of any signs of alien intelligence into a scientific paradox that
changed the very approach to this problem. It became clear that "the silence of the Universe" is a
special phenomenon that must be explained. M. Hart and I.S. Shklovsky proposed a radical
solution to the paradox: THEY are silent, because they simply are not there. Another radical
solution lies in the reply -- whoever was not silent, has ceased to exist -- given by R. Zubrin at a
conference in Colorado during a presentation of one of the authors of this paper. This deep
thought had been expressed before. Let us imagine, for example, that the higher forms of
intelligence related to the lower forms as we do to harmful insects, like locusts. They are
annihilated as soon as the radio picks them up.

The silence of the Universe, however, can also be explained in less radical terms, such as, for
example, that the ages of comparable technological development of civilizations must coincide in
order for them to recognize one another. The characteristic scale of technological development
on Earth, whether you begin with Maxwell-Hertz or with Popov-Marconi, is only 100 or 150
years. Even starting with the ancient Greeks there has elapsed a period of time only one order
greater, whereas the age of the oldest stars is greater by eight orders. We are unable to imagine
the potential of humanity even 100 years from now, not to speak of a billion -- that is, if the
development of science, and of our own existence, will continue that long.

One way or another, it is hard to imagine another scenario than that intelligence first appeared in
the course of biological evolution. Thus we may suppose that during some stage of technical
development and comprehension comparable realities may appear in all civilizations throughout
the Universe. The problem, it seems, lies precisely in the very rapid pace of technological
development, and in the very great disparity, several billion years, in the ages of stars with an
appropriate content of heavy elements, which entails a disparity in the age of planets on which
life may appear, and so of the moments at which intelligent life appears. For this reason, the
probability of finding two civilizations at the same time with comparable levels of development,
permitting one of them to recognize the other, is very small, and the chances of finding intelligent
siblings nearby are negligible.

(Many years ago one of the present authors shared with I.S. Shklovsky his observation, that we
should pay attention to unusual radiation coming from opposite points in the heavens, in the hope
that we may intercept precisely directed radio transmissions between stars. Yet one of the
pioneers in the search for extraterrestrial civilizations became an extreme pessimist around this
time, considering it most likely that we are alone in the Universe. His answer was a sad smile  
that requires, you see, that there be three civilizations on a single line...)

The tiny possibility of encountering intelligent siblings, however, may mean, that we have not
noticed THEM simply because we've hardly looked. Further research is essential, along with
further development of astronomy's tools for observation. A particularly promising avenue is the
study of star clusters, whose stars have roughly the same age, and whose size is measured in
light-weeks, such that the inhabitants of such clusters would be able to conduct practical
exchanges of information. The power of their directed signals may be great enough, and we
might be able to happen upon the extension of a radio beam. Globular clusters of about 15 billion
years in age, containing millions of stars, would be the most interesting, if it weren't for two
circumstances -- their stars contain few heavy elements, and the density of their central parts is so
great that the planets' orbits around stars may be unstable. In the search for life resembling ours,
the best places to look are the old, dispersed clusters with a normal chemical abundance and ages
less than 10 billion years, like M67 and NGC 188.

\section{How Long Has Intelligence Existed?}

There are, however, reasons for remaining suspicious of all these possibilities. We cannot
exclude the possibility that such notions as "life", "technology," and even "intelligence" are
merely the reflections of weakly developed self-organizing cosmic systems. The higher levels of
development are not subject to effective description using the conceptual tools of modern
science. We may recall that B.N. Panovkin insisted on the possibility that different civilizations
can have systems of signs and concepts of radically different types, such that mutual
comprehension is fundamentally impossible.

Let us imagine that technologically and, consequently, scientifically developed civilizations like
those on Earth are aware that in several centuries they will be transformed into objects of a
wholly different kind, in relation to which expressions like "await a response from our intelligent
siblings" have no meaning. In such a case the civilization will have no reason to send signals to
anyone, or to seek to receive them, and this problem simply does not arise. Such a civilization is
like a caterpillar which turns into a butterfly. The caterpillar, if it is aware that it will soon turn
into a butterfly, will not send signals to other caterpillars. It is possible that the comprehension of
such possibilities comes at the same time as the realization of technical means for the
transmission of adequately long-distance signals. Thus we have the silence of the Universe. 

The very notion of intelligence in relation to cosmic civilizations is often understood in too
narrow a sense, as a system's ability to resolve difficult tasks which are essential for survival in a
complex and sometimes hostile environment. In such a context the central question becomes one
of whether intelligence is a characteristic only of biological systems. This problem was posed by
A. Turing in the thirties and carefully considered by the founder of cybernetics, N. Wiener. In
1980, this theme was again sounded by F. Dyson in connection with his analysis of the problems
of immortality, from the point of view of physics. Dyson used a functional definition of life in
whose framework there is an essential structure of relations among elements, yet whose physical
substratum is unimportant. He came to the conclusion that intelligent beings capable of changing
the physical nature of their bodies may potentially exist eternally, but that in the distant future,
when the entropy of the Universe becomes extremely high, they will have to interrupt their
activity for every longer intervals, hibernating and waking up to the sound of a particular "alarm
clocks." Nevertheless, in L. Kraus and G. Starkman (Scientific American, Nov. 1999)
showed that the existence of such an "alarm clock" contradicts the principles of quantum
mechanics. That means that each such intelligent system must eventually go to sleep forever ...

An intelligent system cannot exist eternally, but it can strive for a long life, on the cosmic scale
of time. Our further analysis will touch upon the hypothesis about how highly-developed systems
might resolve this problem.

\section{The Invariants of Perfect Systems}

The most perfect systems that we know are ourselves. Therefore, let us take a look at our "self-
description," as it has been developed in modern psychology. The human being is represented
here in two ways. First of all, it is regarded from the external viewpoint as a particular object.
This permits us to study human behavior and the human brain, in the hope of determing the
objective links between them. The brain, in this aspect, appears as a complex information system.
We must note that in the 1960's holographic means were proposed for the representation of this
system (B. Jules and K. Pennington, P. Westpaik, K. Pribram). The peculiarity of psychological
investigation, however, demands that we study the human being from a fundamentally different,
internal viewpoint. We are interested not only in how the brain works, but also in how the human
being feels this, what pictures appear before the eyes, as it evaluates and realizes itself.
Psychiatry is wholly based not so much on models of the brain, as on sophisticated views of the
inner subjective life. Throughout the 20th century numerous attempts have been made to cross
the barrier separating the inner and outer viewpoints, to create a third, more general one. All
these attempts have been unsuccessful in the end. We apparently must accept the idea, expressed
by Nils Bohr, that the outer and inner views of a phenomenon of consciousness are in a
complementary relationship.

A few years ago one of the authors of this article constructed a formal model of the subject in the
act of choosing between two polar alternatives (V.A. Lefebvre, The Cosmic Subject. Moscow,
Russian Academy of Sciences Institute of Psychology Press, 1997 (in English). Available at
Amazon.). This model's predictions have since passed psychological tests (J. Adams-Webber,
American Journal of Psychology, 1998, V.110, 527-541). The core of the model is a
particular mathematical function, which can be deduced from several elementary propositions
concerning the character of human activity. It turned out that this function possesses one
remarkable characteristic. It can be represented as the repeated composition of a single simpler
function, a representation, moreover, which is unique. This composition was naturally interpreted
as a system of sequential images of the self, possessed by the subject. In other words, the subject
possesses an image of itself, which, has, in turn, an image of itself, and so forth. Each image is a
subject, which exists in the inner world of a certain observing subject. Therefore, the observer
has two complementary points of view: one is external, and the other, is, from the standpoint of
its image, internal. Moreover, it turns out that each image is a "mix" of two states -- positive and
negative.

As a theoretical analogy for the action of the subject this model employs a projection on some
external "screen" of the subject's state. Such a projection can occur either as the creation of an
external self-description, or as self-reproduction (the model does not distinguish between the
two). The further analysis of this model permits us to reveal its formal connection with the
functioning of neural networks, and also with the laws of thermodynamics. It turns out that if
formal neurons in an active state are accidentally connected with other neurons, "infecting" them
by their state, the dynamics of the system may be described by a differential equation whose
solution is a function describing human reflexion (Lefebvre, 1998). In this way a link is
established between reflexion and the dynamics of the neural network. The number of neurons in
this network must be very large, since only in this case is the limit transition justified permitting
us to write a differential equation. Regardless of the neural parallel, it turns out that the
composition of functions describes not only multiple reflexion, but also a chain of abstract heat
machines, which is in essence a particular "paraphrase" of the two first laws of thermodynamics.
The model's logic suggests that the machines are analogues for consistent images of the self, and
that the work they perform is an analogue for subjective experience.

In this way, the model of reflexion turns out to be indissolubly connected with such general
concepts of complex systems as the formal chain of neurons and thermodynamics. This
circumstance inclines us to the thought that the model's mathematical structure is that invariant
which is inherent in all perfect systems, possessing, like the human being, a subjective world and
the ability to multiply reflect it. If we accept this idea, it is natural to suppose that in creating its
external self-description, the system uses the language of that invariant. In this case mathematical
correspondences may constitute one of the aspects of this self-description, connected with the
chain of heat machines. The work performed by these machines, in accord with the formal
model, forms "double" geometrical progressions. This is a sequence of numbers consisting of
two geometrical progressions with the same denominator. One progression corresponds to the
odd members of the initial sequence, the other with the even ones. An example of this double
geometrical progression is the following sequence with the denominator 2:

2, 3, 4, 6, 8, 12, 16, 24

Such sequences permit us to retrieve accurately the subjective state of a system. In this way, the
model brings us to a concrete result. It predicts that in the signals created by perfect systems and
bearing their self-description there may be a double geometrical progression.

\section{Rapid Burster}

We decided to try finding such sequences through analysis of the literature on the best known
sources of alternating x-ray radiation. The broad class of such sources comprises bursters. Each
of these objects, within the framework of our current knowledge, consists of two stars, one
ordinary and one neutron, the latter drawing matter out of its companion. This accumulates on its
surface, leading to periodic thermonuclear explosions which emit x-ray flashes. The idea that x-
ray radiation can be used by intelligent beings for transmitting signals is not new: the American
astronomer R. Corbet, in 1973, wrote that the radiation of bursters could be modulated, for
example, by a metallic screen. These objects are few in number, and we can observe them
throughout our galaxy.

In 1976 the strangest of these objects was discovered in the globular cluster Liller 1, and was
named the Rapid Burster. Aside from the usual thermonuclear flashes, it also produces flashes of
a second kind, whose physical mechanism has remained a mystery; there exists a point of view
that they are connected with the appearance of certain instabilities in the accretion disk. One of
the astonishing features of these flashes of the second type is the lack of correlation between their
form, on the one hand, and their length and intensity, on the other. Such phenomena are very
seldom encountered in non-living systems, though they may be common in artificial
communication systems. For example, the same song may be performed loudly or quietly,
quickly or slowly.

We discovered, that in 1985, the Japanese astronomer I. Tawara and his colleagues, in the
process of analyzing the Rapid Burster's strange flashes, found in them precisely a double
geometrical progression! Such sequences are formed not only by the height of peaks on the
"slope" of their profiles, but also the time intervals between them. It was recently determined,
that "regular" thermonuclear flashes are correlated with a radio source, found a distance of two
and a half parsecs from the Rapid Burster (it is not out of the question, to be sure, that the
disparity of x-ray and radio sources may be due to errors in the x-ray coordinates). Even before
this it had been discovered that the radio source was located almost in the center of the spherical
cluster. Does this all not indicate that the x-ray alternation of Rapid Burster is governed by
intelligent beings living somewhere in that cluster? Alternatively, we may be observing the
natural conscious activity of some intelligent subject of a non-biological character, those
invariant characteristics spoken of earlier? (V.A. Lefebvre and Iu.N. Efremov, Astron. Astroph.
Trans. V. 18, 335-342, 1999).

\section{The Inner Worlds of Black Holes}

The efforts of scientists working to unify the theory of relativity with quantum mechanics are
currently focused on models of black holes -- extremely dense objects, capable of "crushing" the
space around them in such a way that they cannot release either matter, or radiation, or
information. Nearly all researchers are convinced, that black holes exist in the cores of many
galaxies, including our own. The mass of such black holes are millions of times that of the Sun.
Nevertheless, each galaxy must also contain many black holes whose mass is on the order 
of ten times that of the Sun -- these are produced by massive stars as a result of gravitational
collapse following the exhaustion of their nuclear energy resources. Such stars usually rotate
quickly around their axes.

A theoretical model of the collapse of a rotating body was constructed in 1963 by R. Kerr. It
provides a description of the spatio-temporal properties found above the horizon of a black hole.
The purely mathematical model, however, can be analytically continued to an area beyond this
horizon. This area is fundamentally inaccessible to observation by an outside observer. For this
reason, it does not have the status of reality, from a purely instrumental point of view. If we jump
ahead, we will note that this exceptional character of black holes is, at least in one aspect, similar
to the subjective world of the human being. This world, for the external observer, that is, for
another human being, also has no operationally definable status as existent. We still do not know
whether there is an experiment which would enable us to distinguish a being, possessed of
subjectivity, from a soulless but highly refined automaton.

The interpretation of Kerr's model of black holes has shown that this area possesses, for internal
observers, a number of remarkable features. The inside of a black holes is a complex system of
infinite universes containing their own black holes. Within this system it is possible to
distinguish a set of black holes sequentially inserted one within the other. Inside each of these
black holes there is a pair of universes, such that distances within one of them are measured in
positive numbers, and in negative numbers in the other. The components of this paradoxical pair
(from a common-sense viewpoint) can be called the "positive universe" and "negative universe".
These universes are separated by a singularity -- a region of the world in which the density of
matter and the curvature of space are equal to infinity (see J. Gribbin, Unveiling the Edge of
Time, Three Rivers Press, New York, 1992).

The great surprise for physicists working in the field of quantum gravity was the theoretical
conclusion that black holes must evaporate (Ia.B. Zeldovich, A. Starobinsky, S. Hawking). This
conclusion followed on the discovery of connections between black holes and, simultaneously,
both thermodynamics, and quantum mechanics. The second surprise arose, when S. Hawking
suggested the possibility that the absorption of matter and radiation by black holes might lead to
the disappearance of some quantity of information from our world. As we know, the most
general description of a physical system is given by the equations of quantum mechanics. These
equations are reversible, when they relate to micro processes. They allow us, with equal
accuracy, to predict the future states of a quantum system and to retrieve past ones. It turns out,
however, that if there is a black hole in a physical system, the equations describing micro
processes, cannot simultaneously describe both the past and the future of the physical system.
This means, from a formal viewpoint, that there occurs a disappearance of part of the
information.

Further theoretical inquiries, conducted by the Dutch physicist J. t' Hooft, have nevertheless
shown that such a loss is incompatible with the law of conservation of energy. This contradiction,
which has been dubbed the "information paradox," threatened to destroy quantum mechanics,
which is the basis for our modern scientific view of the world (L. Susskind, Black Holes and the
Information Paradox; Scientific American, April 1997). It is therefore no surprise that serious
steps were undertaken to resolve the information paradox. At first it was proposed that the
information absorbed by the black hole might be returned in its process of evaporation. In this
case, however, the emission of particles from various parts of space surrounding the horizon of
the black hole would have to be correlated, and it remains wholly uncertain how such a
correlation might be achieved. A new way of resolving the information paradox was suggested
not long ago by the American physicist L. Susskind, who constructed a holographic model of a
black hole.

Within this model, information does not disappear in the depths of a black hole. It is "deposited"
on its shell or horizon. In other words, the quantity of information on the shell is always precisely
equal to the quantity of information in the substance passing over the horizon. In this way, the
shell of the hole is a particular "text," fixing the complexity of the flow of substance passing over
the horizon. We see that black holes possess certain characteristics of an information system.

Another step toward the understanding of black holes was Susskind's and t' Hooft's
generalization of Nils Bohr's principle of complementarity. The shell of a black hole, together
with the information recorded in it, are a reality for the external observer. An observer, however,
falling freely toward the center of the black hole, discovers no layer of information at the moment
of crossing its threshold. This object, for him, does not have the status of a real existent.
Nevertheless such a distinction of realities for two observers does not present logical difficulties,
since there is no contact possible between them. The observers have no way to counterpose the
results of their individual observations. The fact, that there is no information shell for the internal
observer, does not signify by any means that we cannot regard it as a material object. We can
ascribe the status of an existent to it only because it is real for the external observer, in exactly
the same way as we agree to consider the electron a particle, despite the fact that in many
experiments it behaves like a wave.

Let us recall here that, in accord with Kerr's model, inside the black hole there is found a
sequence of black holes inserted one within the other. Let us now place an observer inside each
of these holes, each capable of observing the subsequent black hole from without. Naturally, the
principle of complementarity is generalized to include such observers. As a result, we arrive at a
series of viewpoints complementarity with respect to one another, connected with the initial
black hole.

\section{Black Holes and the Model of a Reflexive Subject}

Returning from black holes to modern psychology, we can see a striking parallel between the
inner world of a black hole in Kerr's model and the psychological model of a multiply self-aware
subject. In both models there is a sequence of elements one within another, connected with the
positions of observers related by complementarity. Each element contains an unsymmetrical pair
(positive and negative universes in the black hole, positive and negative states in the model of the
subject). In both cases, from the viewpoint of an external observer there exists a thermodynamic
description of the system. Moreover, there are reasons to think that holographic models might
turn out to be effective both for studying the brain and for investigating black holes.

Future research should provide an answer to the question of whether there exist deeper parallels
between the model of black holes and the model of the subject. The establishment of such
parallels might be a substantive step towards a fundamental unification: the inclusion of the
intelligent subject in our physical worldview. Such parallels, however, have not yet been found.
For this reason, our further deliberations on this theme are purely speculative.

\section{A Not-Quite-Scientific Hypothesis}

Nonetheless, we will indulge ourselves in the consideration of a fictive scenario. Since black
holes have "emptiness" for the internal world, they can be turned, by a civilization like that on
Earth, into a gigantic individual, a super-personality, capable of multiple self-awareness and
preserving on its horizon (which may be an analogue of the brain) all the information
accumulated by the civilization.

Let us emphasize that we are speaking of a black hole as of the physical basis of a single gigantic
personality but not as of a new house into which individuals are moving to set up a civilization.
The inner space of a black hole is the subjective world of this individual, which has no reality for
an external observer. Everything found in this subjective world constitutes the "imagination" of
the giant.

Such a "subject," being the successor to biological civilization, may be capable of self-
development and, more importantly, of self-reproduction; in other words, black holes may give
rise to other black holes. Quantum evaporation finally destroys the "intelligent" subject, whose
body is the black hole. Nevertheless, it may exist for an unimaginably long time, even by cosmic
standards. This, in fact, may be the reason that civilizations seek to transform themselves into
such "almost eternal" objects. Yet how the black holes might give birth to a new generation?

According to modern views, for the appearance of a black hole it is essential that the mass of the
collapsing star be at least three or four times that of the Sun. The lifespan of massive stars is
very short on a cosmic scale, only a few million years. In this time a large portion of the nuclear
fuel in the depths of such a star is exhausted. For this reason the light pressure weakens, the
upper levels of the star collapse inward, and there is a huge explosion, called the flash of a
supernova. The star's nucleus is crushed and turned into a black hole.

The actual physical mechanism for the appearance of massive stars is still not wholly understood.
We know only that stars appear as the result of a condensation of interstellar gas. Apparently, in
order for this condensation to produce a massive star, an external force is needed to raise the
pressure in the cloud of gas. The formation region of massive stars is, as a rule, concentrated in
the spiral arms of the galaxy; moreover, in the lifespan of a single galaxy there occur the births
and deaths of many generations of massive stars. The spiral pattern of symmetrical and long arms
is a wave of increased density of stars and gases, which rotates like a solid body, with a constant
angle speed, and on meeting the clouds of gas of the giant disk with already accumulated in the
arms and there is a rise in pressure in the gas, leading ultimately to the formation of massive
stars.

Is it possible to suppose that one of the tasks facing a cosmic subject would be to organize the
"production" of massive stars, which sufficiently quickly, over millions of years, turn into black
holes by natural means? Of course, the process of this creation also must be entirely natural --
there must be some way to initiate the rise in density of the gas clouds. A powerful explosion in
the gas medium is clearly the most economical way to initiate the formation of stars. Such
explosions can be observed, and black holes may be linked in the most direct way with their
appearance. Moreover, we can even speculate that many massive stars -- future black holes -- are
born as a result of the death of a given black hole (following its merging with another component
of the binary system of compact objects) and the appearance of a new black hole. It is believed
that super-powerful explosions occur at the same time, accompanied by flashes of
gamma-radiation.

\section{Star Arcs and Galactic Rings}

Not only a spiral wave of density, but also the flash of a supernova star lead to the appearance of
a spherical wave of raised density gas, in which subsequently young stars take form. This
mechanism is active in irregular galaxies. It has been several decades since the closest of them --
the Large Magellanic Cloud (LMC) revealed strange structures called giant star arcs. They have
been recognized since then in a number of other galaxies. These are huge arc-shaped young star
complexes, with a curvature radius of 200-300 ps. In the centers of these immense star arcs there
are no star clusters containing enough supernovae or 0-stars in them necessary to form arc-
shaped star complexes. The suggestion has been put forward that these arcs are the product of a
superpowerful explosion, creating a spherical wave of density in the cloud of gas. This
hypothesis has been supported by the discovery of gamma-bursts, since they testify that there
may occur even more powerful explosions in the Universe than those of supernovae.

Nevertheless there is one fact, showing that such a simple version of the origin of star arcs,
at least, is incomplete. In the LMC and two other cases the arcs are formed by groups of two to
four arcs. The super-powerful explosions are a fairly rare phenomenon even in cosmic terms.
One might ask, what causes their appearance so close to one another, in only one part of the
galaxy and with an interval of only about 10 million years? One of the authors (Iu.N. Efremov,
Letters to the Astron. J. V. 25, 100-107, 1999; HERALD of Russian Acad. of Sci. Vestnik RAN,
V.70, 314-323, 2000) put forward the hypothesis, that the progenitors of these objects, producing
immense explosions, are born in a dense cluster and ejected from them before they explode.
These are the characteristics to be expected in a binary system of compact objects -- neutron stars
or black holes, whose merging following the transfer of energy from orbital movement into
gravitational wave produces just such gamma-bursts. Numerous massive stars in star arcs born
from superexplosions turn ultimately into black holes.

Yet is the concentration of star arcs produced by superexplosions in close proximity to one
another not the result of the coordinated activity of groups of intelligent black holes? If we accept
this hypothesis, then it may be expected that there exist entire galaxies with the clear signs of the
collective coordination of processes of production of new massive stars, from which black holes
appear as if hatched from the eggs. For some reason their "natural" appearance is not enough, or,
on the other hand, they need to show up precisely in that location, so that measures are taken to
stimulate a local production of massive stars.

We know of an entire galaxy, with features inviting such an explanation -- the Cartwheel. It
consists of a nucleus and two concentric rings. The nucleus and inner ring are formed out of old
stars with little mass, whereas the outer ring consists of bright, young massive stars. From the
internal ring to the outer there are "spokes" (giving the galaxy its name), at the same time these
spokes are directed along tangents to the inner ring.

This galaxy has long attracted the attention of scientists. Pictures taken with the Hubble
Telescope, have permitted a more detailed inspection of fine details in the inner ring. It turns out
that there are objects there resembling comets in form. These objects seem to have a head and a
tail, with the latter in many cases seeming to pass gradually into a spoke. The proposal was put
forward that "comets" these are the traces of a substance falling toward the center of the galaxy.
Nevertheless, such a hypothesis cannot explain the comet-like form of these traces. The
following suspicion arises: are we not observing the remnants of synchronous directed
explosions, whose purpose is to create a ring-shaped wave of density, leading to the formation of
a large number of young massive stars in the outer ring, from which subsequently "ripen" black
holes? If that is so, then, looking at the Cartwheel, we are observing a wave of intelligence,
which was written about a quarter-century ago by Hart and Shklovsky. Yet the possessors of this
intelligence are not biological organisms, but black holes. And the birth of new black holes is
marked by fireworks of gamma-bursts, the most powerful explosions known in all the Universe.

\section{The Presumption of Naturality}

There are, of course, also natural explanations for groups of star arcs and the strange wheel-
shaped galaxy. Near the group of arcs in the LMC there is in fact an old and massive star cluster,
from which compact binary systems with a neutron star or black hole may have escaped, whose
fusion led to the superpowerful explosion and gamma-ray burst. The unusual "Wheel" structure
could have arisen due to the passage through this galaxy's center of another, smaller galaxy,
creating a shock wave, and thus an active process of star formation, appearing to us as the outer
ring.

One way or another, both the star arcs and galaxy with a ring around it are a highly strange
formation, and they may be connected with the activity of a cosmic intelligence, perhaps by some
means as yet completely unfathomable to us... Both "natural" hypotheses raise difficulties. Not
far from the second group of giant star arcs, in the spiral galaxy NGC 6946, there are, it would
seem, no suitable clusters, as will soon be verified by the Hubble Space Telescope. This galaxy
possesses a few other peculiarities as well -- extremely bright or adjacent remnants of
supernovae, the spiral arms of an intensified magnetic field, not coinciding with the arms,
optically visible...

We get the impression that there are galaxies in which the percentage of unusual objects -- star
arcs, rich young clusters or unusual supernovae and their remnants -- is significantly higher than
in others. The intensity of formation of massive stars in them is always elevated, and such objects
are usually connected precisely with a high rate of star formation. This might be explained by
high gas content, but these galaxies do not have a higher relative mass of gas than many others
having no such peculiarities. Does not then NGC 6946 and Cartwheel belong to the number of
galaxies, gripped by the Great Ring of intelligence...

The "presumption of naturality" spoken of by I.S. Shklovsky, the sorely needed attempts to find
"natural" explanations of things and phenomena should not be abolutized or turned into a
prohibition on flights of fantasy. With this thought, expressed in abstract form, everyone is eager
to agree, yet as soon as we refer to a specific object provoking suspicion, the fantasizer begins to
attract strange looks and risks the loss of his scientific reputation... Yet there is no other way than
to keep searching and investigating all strange objects -- always keeping in mind the  possibility
that we will encounter the traces of activity of intelligent subjects (see, for example, N.S.
Kardashev, Voprosy filosofii, 12, 1977).

Let us remember that even artificially created objects and phenomena are unquestionably subject
to the laws of physics, which are definitely known to us in areas remote from singularities. Many
believe that we have long been the witnesses of activity of another intelligence, but do not realize
it. Thus suspicion has fallen on the active cores of galaxies, possessing, like Cyg A, long
narrow jets (sometimes up to several megaparsecs in length), from which are emitted a substance
with subrelative speeds. The collimation mechanism of emissions from quasars and
radiogalaxies, from object SS433 and young stars, -- and possibly, also from objects generating
gamma-bursts -- the mechanism that makes these jets so narrow is not yet completely clear. The
phenomenon is obviously natural, since it is encountered so frequently in nature -- writes the
American astrophysicist B. Paczynsky (Preprint Astro-Phys/9909048, 4 Nov 1999). Yet do these
words not signify that bizarre thoughts occur not only to hopeless crackpots...

\end{document}